\documentclass[prl, twocolumn, longbibliography]{revtex4-1}  
\usepackage{graphicx,graphics,color,epsfig} 
\usepackage{amsmath, amssymb,amsfonts} 
\usepackage{bm,times,xspace,mhchem} 
\usepackage[colorlinks,citecolor=blue,linkcolor=red]{hyperref} 

\begin{document}

\preprint{APS/123-QED}

\title{Experimental Realization of Near-Field Photonic Routing with All-Electric Metasources}

\author{Yang Long$^1$}
\author{Jie Ren$^1$}
\author{Zhiwei Guo$^2$}
 \email{2014guozhiwei@tongji.edu.cn}
\author{Haitao Jiang$^2$}
 \email{jiang-haitao@tongji.edu.cn}
\author{Yuqian Wang$^2$}
\author{Yong Sun$^2$}
\author{Hong Chen$^{1,2}$}
\affiliation{%
$^1$Center for Phononics and Thermal Energy Science, China-EU Joint Center for Nanophononics, Shanghai Key Laboratory of Special Artificial Microstructure Materials and Technology, School of Physics Sciences and Engineering, Tongji University, Shanghai 200092, China\\
$^2$Key Laboratory of Advanced Micro-structure Materials, MOE, School of Physics Science and Engineering, Tongji University, Shanghai, 200092, China
}%

\date{\today}

\begin{abstract}
The spatially confined evanescent microwave photonics have been proved to be highly desirable in broad practical scenarios ranging from robust information communications to efficient quantum interactions. 
However, the feasible applications of these photonics modes are limited due to the lack of fundamental understandings and feasible directional coupling approaches at sub-wavelengths. 
Here, we experimentally demonstrate the efficient near-field photonic routing achieved in waveguides composed of two kinds of single-negative metamaterials. 
Without mimicking the polarization features, we propose all-electric near-field metasource in subwavelength scale and exemplify its near-field functions like Janus, Huygens and spin sources, corresponding to time-reversal, parity-time and parity symmetries of its inner degree of freedom.
Our work furthers the understandings about optical near-field symmetry and feasible engineering approaches of directional couplings, which would pave the way for promising integrated mircrowave photonics devices.
\end{abstract}

\maketitle

Microwave photonics have been attached a lot of attentions for its applications from classical regions to quantum aspects~\cite{marpaung2019integrated,horodynski2019optimal, capmany2007microwave, you2011atomic, gu2017microwave}. The microwave photonics have the millimeter wavelength for flexible on-chip photonics devices~\cite{marpaung2019integrated} and the same energy order for artificial atom physics in superconducting circuits~\cite{you2011atomic, gu2017microwave} or spin cavitroinics~\cite{zhang2014strongly, viennot2015coherent}. 
As result, microwave photonics have become one of the important bridges for communicating information and transferring energy, from the macroscopic signal process devices~\cite{marpaung2019integrated,guo2018enhancement} to the quantum interactions between two artificial atoms~\cite{gu2017microwave}. 
The miniaturized on-chip microwave photonics requires the efficient routing and ultrafast switching for microwave inputs in the deep subwavelength scale~\cite{marpaung2019integrated, atabaki2018integrating}. This is an open challenge due to the lack of understandings about the symmetry and geometry of near-field microwave photonics.

So far, this subwavelength near-field routing is usually achieved by using the local polarizations of light: electric field $\bm{E}$ and magnetic field $\bm{H}$. One well-eatablished way is exploiting the spin-orbit coupling and quantum spin Hall effect (QSHE) states of light~\cite{bliokh2015quantum, lodahl2017chiral, aiello2015transverse, bliokh2015spin, sollner2015deterministic}. The particle scatterings of the incident chiral light~\cite{bliokh2015quantum} or the local chiral electric dipoles~\cite{rodriguez2013near} will excite the directional surface wave due to the non-zero transverse spin ($\propto{\rm Im}[\bm{E}^*\times\bm{E} + \bm{H}^*\times\bm{H}]$) and spin-momentum locking~\cite{bliokh2015quantum, van2016universal}, smiliar counterparts of which can be found universally in other wave systems, such as acoustic~\cite{shi2019observation} and elastic waves~\cite{long2018intrinsic}. 
The other way is based on the relative phase delay combinations of electric and magnetic fields~\cite{picardi2018janus, picardi2019experimental, F.Picardi2019}. The super-positions of geometrically orthogonal electric and magnetic dipoles can excite the surface waves associated with Poynting vectors ($\propto{\rm Re}[\bm{E}^*\times\bm{H}]$) and reactive power ($\propto{\rm Im}[\bm{E}^*\times\bm{H}]$) respectively, which result in different behaviours beyond spin-momentum locking~\cite{picardi2018janus,picardi2019experimental, F.Picardi2019}. 
These approaches provide good understandings about the near-field photonics routing.
It implies that the symmetry of near-field excitations would be  associated with that of sources. 
The question will be whether we can achieve the photonics routing from a deep perspective, such as  the symmetry features of near-field photonic systems.

\begin{figure*}[tp!]
\centering
\includegraphics[width=\linewidth]{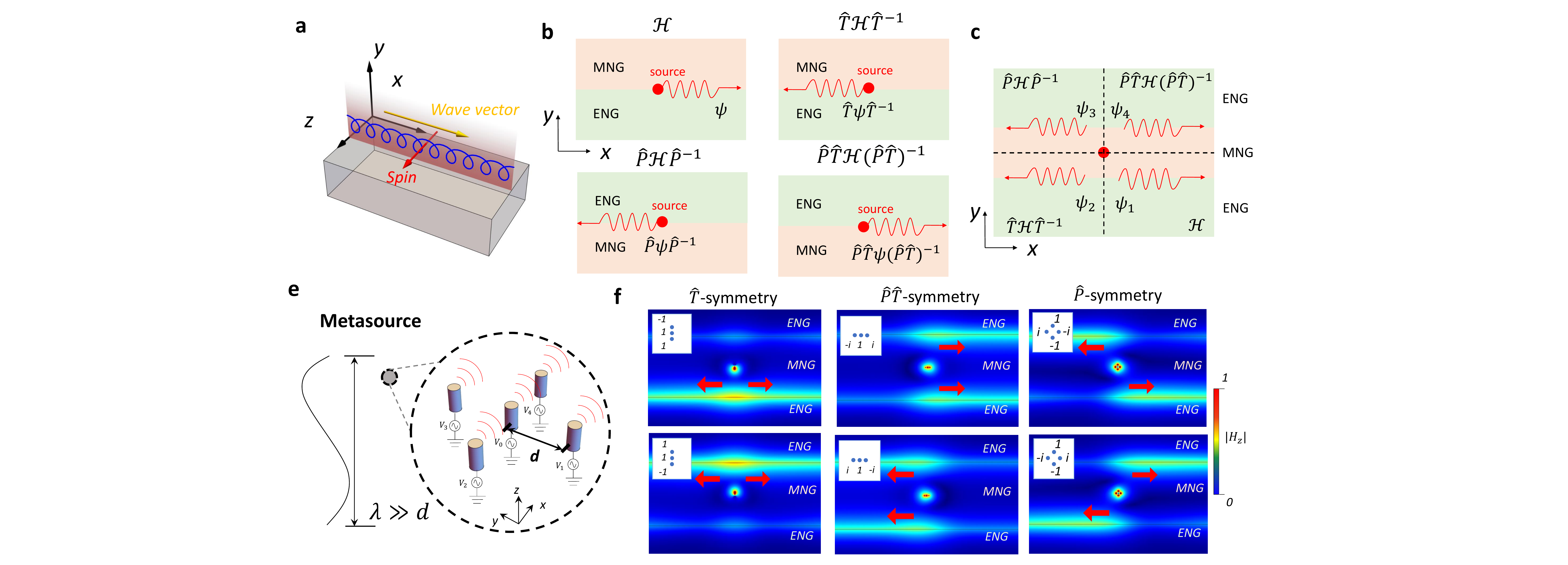}
\caption{The near-field photonic routing of all-electric metasources based on symmetry analysis. \textbf{a}, the dynamic properties of the evanescent photonic mode. \textbf{b}, the photonics planar waveguide system $\mathcal{H}$ composed of the epsilon-negative (ENG, namely $\varepsilon < 0$ and $\mu >0$) and mu-negative (MNG, namely $\varepsilon > 0$ and $\mu <0$) meta-material can support the non-trivial QSHE state. \textbf{c}, the combined system that contains the four different symmetry parts with respect to the source. \textbf{e}, the metasource is composed of electric sources and its geometrical scale is smaller than the working wavelength, namely in deep subwavelength. \textbf{f}, the demonstrations of near-field routing behaviours of metasources with different symmetries.  
The normalized $|H_z|$ fields are plotted.
The height of central MNG is $30$cm, the working frequency is $2.7$ GHz. $d=1$ mm ($d \ll \lambda$). $\varepsilon=-9.44$, $\mu = 1.00$ for ENG, $\varepsilon=6.58$, $\mu=-0.35$ for MNG.~\cite{guo2017photonic}}
\label{fig:symmetry}
\end{figure*}

In this Letter, we demonstrate that the efficient near-field microwave photonics routing based on the structured all-electric metasource in the subwavelength scale. The metasource is composed of only the phase-delayed electric source elements, such as voltage ports or electric dipoles, which will be placed in the subwavelength area. 
It is known that the magnetic dipole with strong strength (c factor ~\cite{picardi2018janus}  compared with electric dipole) is usually hard to achieve practically due to weak interactions between almost optical media and magnetic fields, and design complexities in broadband magnetic meta-materials. 
Different from the schemes based on the polarization profiles~\cite{bliokh2015quantum, rodriguez2013near, picardi2018janus, picardi2019experimental, F.Picardi2019}, the physical mechanism underlying our method is the symmetry transformation invariance between the environments and the inner degree of freedom in metasources.
In our work, three kinds of symmetries: parity ($\hat{P}: \bm{r}\rightarrow -\bm{r}$), time-reversal ($\hat{T}:t\rightarrow -t$) and parity-time ($\hat{P}\hat{T}: \bm{r}\rightarrow -\bm{r}, t\rightarrow -t$) symmetry, have been discussed. 
We found that the metasource can reproduce effective Janus source under $\hat{T}$-symmetry, Huygens source under $\hat{P}\hat{T}$-symmetry and spin source under $\hat{P}$-symmetry. 
We experimentally verify these predicted phenomena based on the 2D microwave photonic system experimentally.

The surface evanescent wave can possess some special dynamic properties, such as supermomentum~\cite{huard1978measurement,matsudo1998pseudomomentum}, chiral polarized field and transverse spin, as shown in Fig.~\ref{fig:symmetry}(a). They are the core essences for the polarization-based schemes~\cite{bliokh2015quantum, rodriguez2013near, picardi2018janus, picardi2019experimental, F.Picardi2019}. Here, we consider one interesting QSHE state that happens on the interface between two different metamaterials~\cite{guo2017photonic}: epsilon-negative (ENG) and mu-negative (MNG), which correspond to two types of topological origins and could be described by the complex Chern number for optical helicity~\cite{bliokh2019topological}. For simplicity, we focus on the transverse magnetic (TM) QSHE mode.  As shown in Fig.~\ref{fig:symmetry}(b), we consider the surface wave mode $\psi$ is supported by the photonics system $\mathcal{H}$ and excited by the harmonic oscillated source term $\psi_s$ with the frequency $\omega$, which can be represented as the Schr\"odinger equation like form:
\begin{equation}
\omega \psi = \mathcal{H} \psi + \psi_s.
\end{equation}
Its symmetry analysis will be: (1) If we apply the parity transformation $\hat{P}$, the equation will be:
\begin{equation}
\omega \hat{P}\psi\hat{P}^{-1} =  \hat{P}\mathcal{H} \hat{P}^{-1}  \hat{P}\psi \hat{P}^{-1}  +  \hat{P}\psi_s \hat{P}^{-1},
\end{equation}
which reflects a fact that: if the $\psi_s =  \hat{P}\psi_s \hat{P}^{-1}$, the source will excite the $\psi$ in the system $\mathcal{H}$ and $\hat{P}\psi \hat{P}^{-1}$ in the system $\hat{P}\mathcal{H} \hat{P}^{-1}$ simultaneously; (2) If we apply the time-reversal transformation $\hat{T}$:
\begin{equation}
\omega \hat{T}\psi \hat{T}^{-1} =  \hat{T}\mathcal{H} \hat{T}^{-1}  \hat{T}\psi\hat{T}^{-1}  +  \hat{T}\psi_s \hat{T}^{-1} 
\end{equation}
which reflects a fact that: if the $\psi_s =  \hat{T}\psi_s \hat{T}^{-1}$, the source will excite the $\psi$ in the system $\mathcal{H}$ and $\hat{T}\psi\hat{T}^{-1}$ in the system $\hat{T}\mathcal{H} \hat{T}^{-1}$ simultaneously; (3) If we apply the parity-time transformation $\hat{P}\hat{T}$:
\begin{equation}
\begin{aligned}
\omega \hat{P}\hat{T}\psi (\hat{P}\hat{T})^{-1} &= \hat{P}\hat{T}\mathcal{H}(\hat{P}\hat{T})^{-1} \hat{P}\hat{T}\psi (\hat{P}\hat{T})^{-1} \\
&+  \hat{P}\hat{T}\psi_s (\hat{P}\hat{T})^{-1}
\end{aligned}
\end{equation}
which reflects a fact that: if the $\psi_s =  \hat{P}\hat{T}\psi_s (\hat{P}\hat{T})^{-1}$, the source will excite the $\psi$ in the system $\mathcal{H}$ and $ \hat{P}\hat{T}\psi(\hat{P}\hat{T})^{-1}$ in the system $ \hat{P}\hat{T}\mathcal{H} ( \hat{P}\hat{T})^{-1}$ simultaneously. These analysis is the core point of this work: the directional routing can be realized by the symmetry properties of $\mathcal{H}$ but without considering the polarization details in $\psi$. 
Taking these facts together, we can construct such a system in Fig.~\ref{fig:symmetry}(c) and there will exist four modes $\{\psi_i\}$ ($i=1..4$) corresponding to the four systems $\mathcal{H}$, $\hat{T}\mathcal{H}\hat{T}^{-1}$, $\hat{P}\mathcal{H}\hat{P}^{-1}$ and $\hat{P}\hat{T}\mathcal{H}(\hat{P}\hat{T})^{-1}$. It's obvious that $\hat{P}(\psi_1 + \psi_3)\hat{P}^{-1} = \psi_1 + \psi_3$, $\hat{T}(\psi_1 + \psi_2)\hat{T}^{-1} = \psi_1 + \psi_2$, $\hat{P}\hat{T}(\psi_1 + \psi_4)(\hat{P}\hat{T})^{-1}= \psi_1 + \psi_4$.

\begin{figure}[tp!]
\centering
\includegraphics[width=\linewidth]{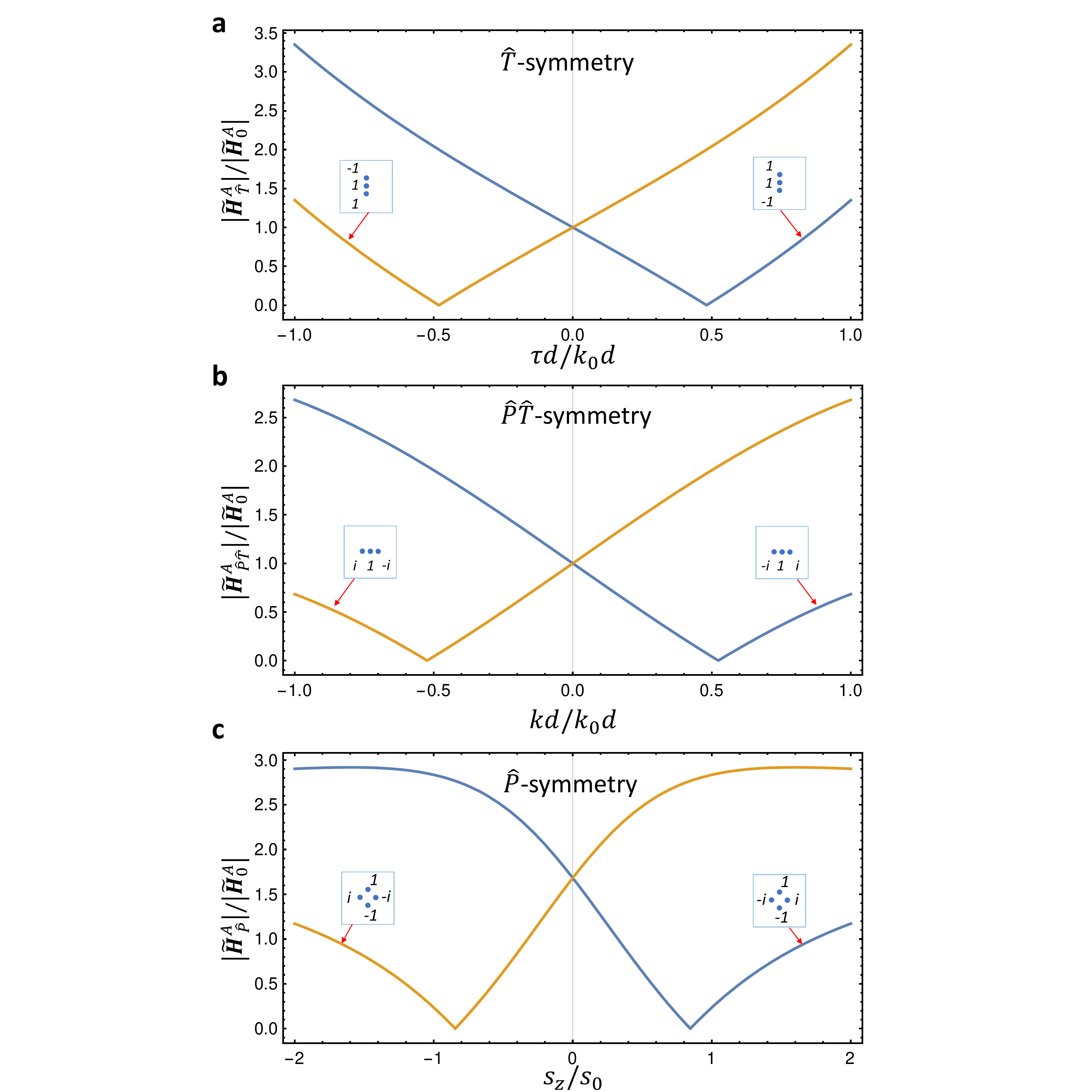}
\caption{Magnetic field amplitude angular spectra of  metasources in a homogeneous medium as a function of $\tau$, $k$ and spin $s_z$ in the $xOy$ plane. \textbf{a}, $\hat{T}$-symmetry metasource (Janus source). \textbf{b}, $\hat{P}\hat{T}$-symmetry metasource (Huygens source). \textbf{c}, $\hat{P}$-symmetry metasource (Spin source). Here, $\widetilde{\bm{H}}^A_0 = \frac{\omega k_0}{8\pi^2 c \tau}$ and $\bm{s}_0$ is the spin angular momentum when $k = \tau$, $\bm{s}_0 = s_0 \bm{e}_z$.}
\label{fig:angularspectrum}
\end{figure}

For achieving a source $\psi_s$ that satisfies the symmetry conditions, we propose a near-field meta-source composed of only electric voltage sources $\{\bm{v}_j\}$, shown in Fig.~\ref{fig:symmetry}(e). The meta-source is of subwavelength scale and constructed by five voltage sources. Their amplitude and phase can be modulated independently, $i.e$, 1 means the normalized amplitude, $i$ means the normalized amplitude with $\pi/2$ phase delay and $0$ (not shown) means the zero amplitude or to be turned off.  According to the above symmetry analysis, we can propose three sources as shown in the insets of Fig.~\ref{fig:symmetry}(f): (1) $\hat{T}$-symmetry will lead to  the mode pairs excitations $\{\psi_1,\psi_2\}$ or $\{\psi_3,\psi_4\}$;  (2)$\hat{P}\hat{T}$-symmetry will be responsible for the mode pairs $\{\psi_1,\psi_4\}$ or $\{\psi_2,\psi_3\}$;  (3)$\hat{P}$-symmetry case excites the mode pairs $\{\psi_1,\psi_3\}$ or $\{\psi_2,\psi_4\}$. From the simulations shown in Fig.~\ref{fig:symmetry}(f), we can see that the same near-field behaviours as the previously proposed Janus, Huygens and spin sources~\cite{picardi2018janus} are reproduced but without magnetic dipoles, not relying on the field polarization $\bm{E}$ and $\bm{H}$. 

\begin{figure}[tp!]
\centering
\includegraphics[width=\linewidth]{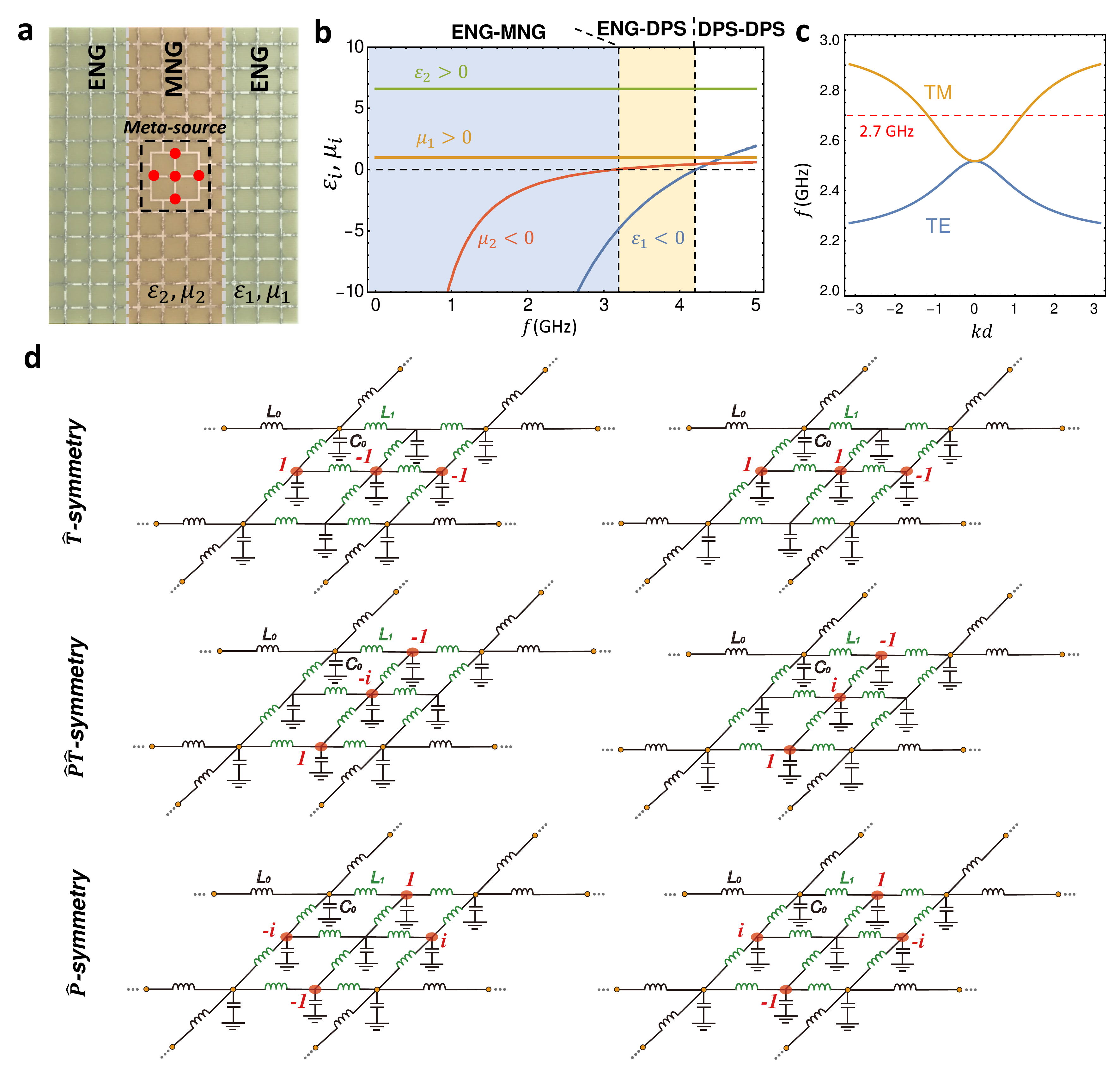}
\caption{Microwave transmission lines metamaterial set-up. \textbf{a}, the photograph of the TLs samples. The sample contains an effective ENG metamaterial loaded with shunted lumped inductors and an effective MNG metamaterial loaded with series lumped capacitors. \textbf{b}, effective parameters of two kinds of meta-materials. DPS means double positive. \textbf{c}, dispersion relations of guided QSHE modes in ENG/MNG waveguides. \textbf{d}, the realization of metasource in microwave photonics TLs system. The voltage ports (Red point) are fed by the unit amplitude but with different phase delays. The working frequency is chosen as 2.7 GHz.}
\label{fig:source}
\end{figure}

\begin{figure*}[tp!]
\centering
\includegraphics[width=\linewidth]{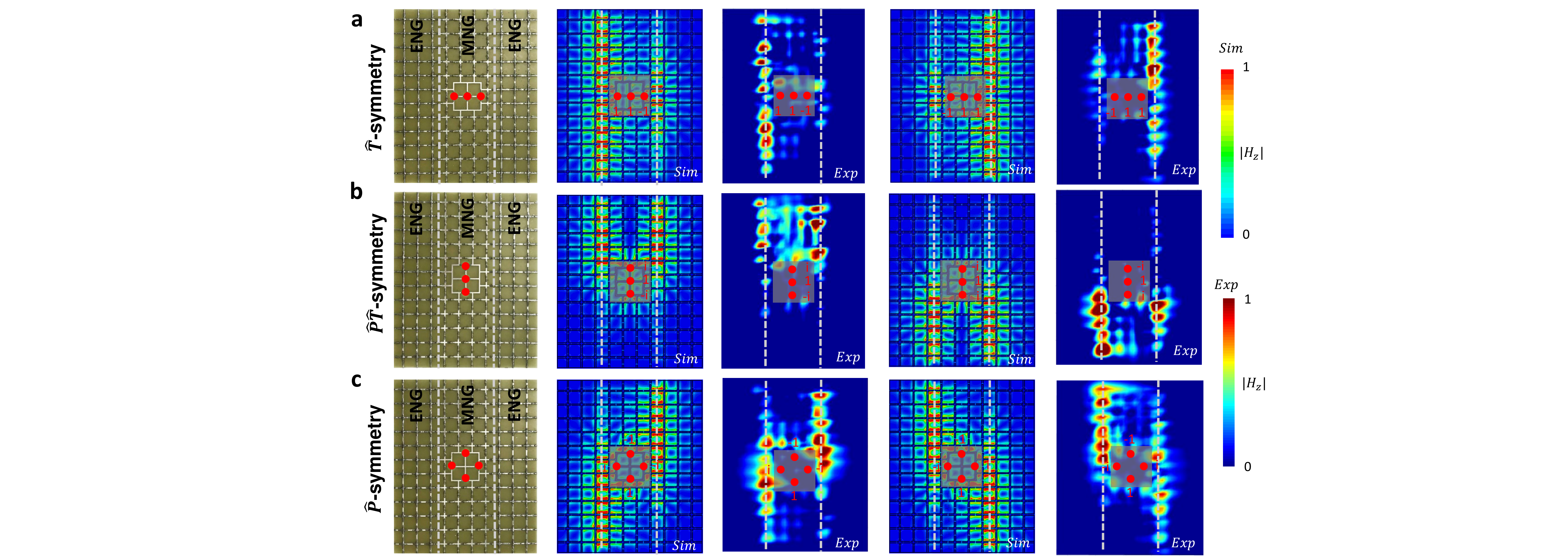}
\caption{Experimental verifications for all-electric metasource based on 2D microwave TL photonics systems. \textbf{a}, $\hat{T}$-symmetry. \textbf{b}, $\hat{P}\hat{T}$-symmetry. \textbf{c}, $\hat{P}$-symmetry. The normalized simulated and the measured normalized $|H_z|$ patterns have been plotted. The settings of the voltage ports have been shown in the figures.}
\label{fig:exp}
\end{figure*}

Besides the symmetry analysis, we will give the angular spectrum descriptions about their near-field behaviour. The electric source in metasource is the voltage source, which can be modeled as magnetic current or magnetic frill generator~\cite{balanis1999advanced}. Considering the harmonic oscillation magnetic current $\mathfrak{M}$ in the frequency $\omega$, an effective magnetic dipole can be associated with an magnetic current density  $\mathfrak{M} = -i\omega\mu\delta^3(\bm{r}-\bm{r}_0)\bm{m}$, where $\bm{m}$ is the dipole moment and $\delta(\bm{r})$ is the delta function. This effective magnetic dipole is originated from the mathematical equivalence of the model trick but not the real one. The magnetic field induced by magnetic dipole array in a homogeneous medium can be written as $\bm{H}^A = k_0^2\bm{\Pi}_m + \nabla(\nabla\cdot\bm{\Pi}_m)$, $k_0^2 = \varepsilon \mu \frac{\omega^2}{c^2}$ and the magnetic Hertz potential $\bm{\Pi}_m$ can be represented as the Green's function form~\cite{picardi2017unidirectional,PhysRevApplied.12.024065}:
\begin{equation}
\bm{\Pi}_m = \sum_j \frac{\bm{m}_j}{4\pi}\frac{e^{i k|\bm{r}-\bm{r}_j|}}{|\bm{r}-\bm{r}_j|}.
\end{equation}
Based on the Weyl's identity~\cite{ishimaru2017electromagnetic}, we can obtain the angular spectrum of $\bm{H}^A$ as:
\begin{equation}
\begin{aligned}
&\widetilde{\bm{H}}^A(k_x,k_y) = \frac{i\omega}{8\pi^2}\frac{k_0}{k_y}\frac{1}{c} \\
&\times\sum_j ((\bm{e}_s \cdot\bm{m}_j)\bm{e}_s + (\bm{e}^{\pm}_{p}\cdot\bm{m}_j)\bm{e}^{\pm}_{p}) e^{-i (k_x x_j + k_y y_j)}
\end{aligned}
\end{equation}
where $\bm{e}_s(k_x,k_y)$ and $\bm{e}^{\pm}_{p}(k_x,k_y)$ are the unit polarization vector for $s$-polarized and $p$-polarized fields~\cite{picardi2017unidirectional}, $k_x^2+k_y^2 = k_0^2$. Here, we consider that 2D photonics case ($k_z =0$) and the magnetic current is out of $xOy$ plane, which means that all effective magnetic dipoles polarized along the $z$ direction, namely $\bm{m}_j = (0,0,f_j)^T$. The angular spectrum will be represented as:
\begin{equation}
\widetilde{\bm{H}}^A(k_x,k_y) = \frac{i\omega}{8\pi^2}\frac{k_0}{k_y}\frac{1}{c} \sum_j f_j \bm{e}_s e^{-i (k_x x_j + k_y y_j)}.
\label{eq:AMspectrum}
\end{equation}
where $\bm{e}_s = \frac{k_x}{k_0} \bm{e}_z$.

Here, we will give the theoretical details in Fig.~\ref{fig:symmetry}(f) based on Eq.~\ref{eq:AMspectrum}. For $\hat{T}$-symmetry metasource(Janus source), there are three voltage sources: $\mathfrak{M}=1 V$ at the center position $(0,0)$, $\mathfrak{M}=\mp 1 V$ at the position $(0,d)$ and $\mathfrak{M}=\pm 1 V$ at the position $(0,-d)$. The evanescent photonics modes can be represented with the complex wave vector $\bm{k} = (k, i\tau,0)$. The angular spectrum for $\hat{T}$-symmetry will be:
\begin{equation}
\widetilde{\bm{H}}^A_{\hat{T}} (k, \tau) = \frac{\omega}{8\pi^2}\frac{k_0}{\tau}\frac{1}{c} (1 \pm e^{-\tau d} \mp e^{\tau d}) \bm{e}_s.
\label{eq:AM_T}
\end{equation}
When $\tau d \ll 1$,  $|\widetilde{\bm{H}}^A_{\hat{T}}|\propto |1\mp 2\tau d|$ , which reflects that the $\hat{T}$-symmetry meta-source will be strongly associated with the decay rate $\tau$, as shown in Fig.~\ref{fig:angularspectrum}(a). In the same process, the $\hat{P}\hat{T}$-symmetry metasource (Huygens source), it will be:
\begin{equation}
\widetilde{\bm{H}}^A_{\hat{P}\hat{T}} (k, \tau) = \frac{\omega}{8\pi^2}\frac{k_0}{\tau}\frac{1}{c} (1 \pm i e^{-i k d} \mp i e^{i k d}) \bm{e}_s
\label{eq:AM_PT}
\end{equation}
and $|\widetilde{\bm{H}}^A_{\hat{P}\hat{T}}|\propto |1 \pm 2 k d|$ when $kd \ll 1$, which shows that the $\hat{P}\hat{T}$-symmetry will result in the $k$-dependent coupling strength, as shown in Fig.~\ref{fig:angularspectrum}(b). For $\hat{P}$-symmetry metasource(spin source), it will be:
\begin{equation}
\widetilde{\bm{H}}^A_{\hat{P}} (k, \tau) = \frac{\omega}{8\pi^2}\frac{k_0}{\tau}\frac{1}{c} (e^{-\tau d} - e^{\tau d} \pm i e^{-i k d} \mp i e^{i k d}) \bm{e}_s
\label{eq:AM_P}
\end{equation}
and $|\widetilde{\bm{H}}^A_{\hat{P}}|\propto 2d|\tau\pm k|$ when $d\ll\lambda$, which reflects that the meta-source with $\hat{P}$-symmetry will be strongly locked with the spin angular momentum $s_z \propto \tau k$, as shown in Fig.~\ref{fig:angularspectrum}(c). In the following, we will verify the near-field behaviour of meta-source experimentally.

In microwave photonics regime, the two-dimensional (2D) transmission lines (TLs) loaded with lumped circuit elements will be a convenient and simple platform to realize arbitrary effective $\varepsilon$ and $\mu$ and observe optical wave propagations. Until now, many high-performance metamaterials have been constructed to achieve the required optical responses in this platform and enable extensive applications, such as cloaking~\cite{PhysRevApplied.10.064048}, hyperbolic dispersion~\cite{kapitanova2014photonic} and topological photonics~\cite{guo2017photonic,li2018topological}.  Here, we design a TL system in Fig.~\ref{fig:source}(a), which achieve the equivalent photonic system in Fig.~\ref{fig:symmetry}(c). The effective parameters can be derived according to the TLs effective medium theory, which have been calculated in Fig.~\ref{fig:source}(b). We can see that the TL metamaterials will support TM modes in the frequency range $2.3\sim3.0$ GHz as Fig.~\ref{fig:source}(c). As shown in Fig.~\ref{fig:source}(d), the voltage port arrays with spatial distributed amplitude and phase have been used  to induce effective near-field sources. The simulated and experimental results of these sources are shown in Fig.~\ref{fig:exp}. The effective Janus source will excite the mode pairs with $\hat{T}$-symmetry, the Huygens source will induce the branches with $\hat{P}\hat{T}$-symmetry and the spin source will stimulate the branches with $\hat{P}$-symmetry. The experimental results are in good agreement with the theoretical analysis and numerical simulations. (More details in Supplementary)

Finally, we discuss the physical reason for the same behaviour between our symmetry-based scheme and the polarization proposals in the previous studies~\cite{bliokh2015quantum, rodriguez2013near, picardi2018janus, picardi2019experimental}. 
The main reason is that we share the same symmetry features as the near field physical quantities: Poynting vector $\bm{\mathcal{J}}$, reactive power $\bm{\mathcal{R}}$ and spin density $\bm{s}$. One can find that: $\bm{\mathcal{J}}$ satisfies $\hat{P}\hat{T}$-symmetry ($\hat{P}\hat{T}\bm{\mathcal{J}}(\hat{P}\hat{T})^{-1}=\bm{\mathcal{J}}$), $\bm{\mathcal{R}}$ satisfies $\hat{T}$-symmetry ($\hat{T}\bm{\mathcal{R}}\hat{T}^{-1}=\bm{\mathcal{R}}$) and $\bm{s}$ satisfies $\hat{P}$-symmetry ($\hat{P}\bm{s}\hat{P}^{-1}=\bm{s}$). The original Janus, Huygens and spin sources are strictly associated with these physical quantities and thus will inherit these symmetry properties naturally. Our scheme removed the requirement of magnetic dipoles and become experimentally tractable. For example, in the optical regime, the meta-source can be realized by the nano-particle arrays or other scatters in the subwavelength scale. 

To summarize, we have proposed the subwavelength all-electric meta-sources for the near-field microwave photonics routing based on parity, time-reversal, parity-time symmetry. 
According to these symmetry features, we have exemplified and observed the fertile functions of excitations supported on MNG/ENG interface based on photonic microwave TL system, corresponding to the Janus, Huygens and spin sources. 
We have shown alternative approach to design the near-field photonics sources  in additions to the polarization engineering, which reflects the inner symmetry properties of the near-field systems.
Our work would improve the understanding about the geometry and topology in near field and inspire new ideas for controlling photonic evanescent modes, i.e, selective wireless energy transfer~\cite{assawaworrarit2017robust, kurs2007wireless} and future integrated optical devices~\cite{gong2018nanoscale, neugebauer2019emission, wang2016directional}.

\begin{acknowledgments}
This work is supported by the National Key R\&D Program of China (Grant No. 2016YFA0301101), by the National Natural Science Foundation of China (NSFC) (Grants No. 11775159, and No. 61621001), by the Natural Science Foundation of Shanghai (Grants No. 18ZR1442800, No. 17ZR1443800, No. 18JC1410900), by China Postdoctoral Science Foundation (Grants No. 2019TQ0232, No. 2019M661605), and by the Opening Project of Shanghai Key Laboratory of Special Artificial Microstructure Materials and Technology.
\end{acknowledgments}

\bibliography{references}

\begin{thebibliography}{37}%
\makeatletter
\providecommand \@ifxundefined [1]{%
 \@ifx{#1\undefined}
}%
\providecommand \@ifnum [1]{%
 \ifnum #1\expandafter \@firstoftwo
 \else \expandafter \@secondoftwo
 \fi
}%
\providecommand \@ifx [1]{%
 \ifx #1\expandafter \@firstoftwo
 \else \expandafter \@secondoftwo
 \fi
}%
\providecommand \natexlab [1]{#1}%
\providecommand \enquote  [1]{``#1''}%
\providecommand \bibnamefont  [1]{#1}%
\providecommand \bibfnamefont [1]{#1}%
\providecommand \citenamefont [1]{#1}%
\providecommand \href@noop [0]{\@secondoftwo}%
\providecommand \href [0]{\begingroup \@sanitize@url \@href}%
\providecommand \@href[1]{\@@startlink{#1}\@@href}%
\providecommand \@@href[1]{\endgroup#1\@@endlink}%
\providecommand \@sanitize@url [0]{\catcode `\\12\catcode `\$12\catcode
  `\&12\catcode `\#12\catcode `\^12\catcode `\_12\catcode `\%12\relax}%
\providecommand \@@startlink[1]{}%
\providecommand \@@endlink[0]{}%
\providecommand \url  [0]{\begingroup\@sanitize@url \@url }%
\providecommand \@url [1]{\endgroup\@href {#1}{\urlprefix }}%
\providecommand \urlprefix  [0]{URL }%
\providecommand \Eprint [0]{\href }%
\providecommand \doibase [0]{http://dx.doi.org/}%
\providecommand \selectlanguage [0]{\@gobble}%
\providecommand \bibinfo  [0]{\@secondoftwo}%
\providecommand \bibfield  [0]{\@secondoftwo}%
\providecommand \translation [1]{[#1]}%
\providecommand \BibitemOpen [0]{}%
\providecommand \bibitemStop [0]{}%
\providecommand \bibitemNoStop [0]{.\EOS\space}%
\providecommand \EOS [0]{\spacefactor3000\relax}%
\providecommand \BibitemShut  [1]{\csname bibitem#1\endcsname}%
\let\auto@bib@innerbib\@empty
\bibitem [{\citenamefont {Marpaung}\ \emph {et~al.}(2019)\citenamefont
  {Marpaung}, \citenamefont {Yao},\ and\ \citenamefont
  {Capmany}}]{marpaung2019integrated}%
  \BibitemOpen
  \bibfield  {author} {\bibinfo {author} {\bibfnamefont {David}\ \bibnamefont
  {Marpaung}}, \bibinfo {author} {\bibfnamefont {Jianping}\ \bibnamefont
  {Yao}}, \ and\ \bibinfo {author} {\bibfnamefont {Jos{\'e}}\ \bibnamefont
  {Capmany}},\ }\bibfield  {title} {\enquote {\bibinfo {title} {Integrated
  microwave photonics},}\ }\href@noop {} {\bibfield  {journal} {\bibinfo
  {journal} {Nat. Photonics}\ }\textbf {\bibinfo {volume} {13}},\ \bibinfo
  {pages} {80--90} (\bibinfo {year} {2019})}\BibitemShut {NoStop}%
\bibitem [{\citenamefont {Horodynski}\ \emph {et~al.}(2019)\citenamefont
  {Horodynski}, \citenamefont {K{\"u}hmayer}, \citenamefont {Brandst{\"o}tter},
  \citenamefont {Pichler}, \citenamefont {Fyodorov}, \citenamefont {Kuhl},\
  and\ \citenamefont {Rotter}}]{horodynski2019optimal}%
  \BibitemOpen
  \bibfield  {author} {\bibinfo {author} {\bibfnamefont {Michael}\ \bibnamefont
  {Horodynski}}, \bibinfo {author} {\bibfnamefont {Matthias}\ \bibnamefont
  {K{\"u}hmayer}}, \bibinfo {author} {\bibfnamefont {Andre}\ \bibnamefont
  {Brandst{\"o}tter}}, \bibinfo {author} {\bibfnamefont {Kevin}\ \bibnamefont
  {Pichler}}, \bibinfo {author} {\bibfnamefont {Yan~V}\ \bibnamefont
  {Fyodorov}}, \bibinfo {author} {\bibfnamefont {Ulrich}\ \bibnamefont {Kuhl}},
  \ and\ \bibinfo {author} {\bibfnamefont {Stefan}\ \bibnamefont {Rotter}},\
  }\bibfield  {title} {\enquote {\bibinfo {title} {Optimal wave fields for
  micromanipulation in complex scattering environments},}\ }\href@noop {}
  {\bibfield  {journal} {\bibinfo  {journal} {Nat. Photonics}\ ,\ \bibinfo
  {pages} {1--5}} (\bibinfo {year} {2019})}\BibitemShut {NoStop}%
\bibitem [{\citenamefont {Capmany}\ and\ \citenamefont
  {Novak}(2007)}]{capmany2007microwave}%
  \BibitemOpen
  \bibfield  {author} {\bibinfo {author} {\bibfnamefont {Jos{\'e}}\
  \bibnamefont {Capmany}}\ and\ \bibinfo {author} {\bibfnamefont {Dalma}\
  \bibnamefont {Novak}},\ }\bibfield  {title} {\enquote {\bibinfo {title}
  {Microwave photonics combines two worlds},}\ }\href@noop {} {\bibfield
  {journal} {\bibinfo  {journal} {Nat. Photonics}\ }\textbf {\bibinfo {volume}
  {1}},\ \bibinfo {pages} {319} (\bibinfo {year} {2007})}\BibitemShut {NoStop}%
\bibitem [{\citenamefont {You}\ and\ \citenamefont
  {Nori}(2011)}]{you2011atomic}%
  \BibitemOpen
  \bibfield  {author} {\bibinfo {author} {\bibfnamefont {JQ}~\bibnamefont
  {You}}\ and\ \bibinfo {author} {\bibfnamefont {Franco}\ \bibnamefont
  {Nori}},\ }\bibfield  {title} {\enquote {\bibinfo {title} {Atomic physics and
  quantum optics using superconducting circuits},}\ }\href@noop {} {\bibfield
  {journal} {\bibinfo  {journal} {Nature}\ }\textbf {\bibinfo {volume} {474}},\
  \bibinfo {pages} {589} (\bibinfo {year} {2011})}\BibitemShut {NoStop}%
\bibitem [{\citenamefont {Gu}\ \emph {et~al.}(2017)\citenamefont {Gu},
  \citenamefont {Kockum}, \citenamefont {Miranowicz}, \citenamefont {Liu},\
  and\ \citenamefont {Nori}}]{gu2017microwave}%
  \BibitemOpen
  \bibfield  {author} {\bibinfo {author} {\bibfnamefont {Xiu}\ \bibnamefont
  {Gu}}, \bibinfo {author} {\bibfnamefont {Anton~Frisk}\ \bibnamefont
  {Kockum}}, \bibinfo {author} {\bibfnamefont {Adam}\ \bibnamefont
  {Miranowicz}}, \bibinfo {author} {\bibfnamefont {Yu-xi}\ \bibnamefont {Liu}},
  \ and\ \bibinfo {author} {\bibfnamefont {Franco}\ \bibnamefont {Nori}},\
  }\bibfield  {title} {\enquote {\bibinfo {title} {Microwave photonics with
  superconducting quantum circuits},}\ }\href@noop {} {\bibfield  {journal}
  {\bibinfo  {journal} {Phys. Rep.}\ }\textbf {\bibinfo {volume} {718}},\
  \bibinfo {pages} {1--102} (\bibinfo {year} {2017})}\BibitemShut {NoStop}%
\bibitem [{\citenamefont {Zhang}\ \emph {et~al.}(2014)\citenamefont {Zhang},
  \citenamefont {Zou}, \citenamefont {Jiang},\ and\ \citenamefont
  {Tang}}]{zhang2014strongly}%
  \BibitemOpen
  \bibfield  {author} {\bibinfo {author} {\bibfnamefont {Xufeng}\ \bibnamefont
  {Zhang}}, \bibinfo {author} {\bibfnamefont {Chang-Ling}\ \bibnamefont {Zou}},
  \bibinfo {author} {\bibfnamefont {Liang}\ \bibnamefont {Jiang}}, \ and\
  \bibinfo {author} {\bibfnamefont {Hong~X}\ \bibnamefont {Tang}},\ }\bibfield
  {title} {\enquote {\bibinfo {title} {Strongly coupled magnons and cavity
  microwave photons},}\ }\href@noop {} {\bibfield  {journal} {\bibinfo
  {journal} {Phys. Rev. Lett.}\ }\textbf {\bibinfo {volume} {113}},\ \bibinfo
  {pages} {156401} (\bibinfo {year} {2014})}\BibitemShut {NoStop}%
\bibitem [{\citenamefont {Viennot}\ \emph {et~al.}(2015)\citenamefont
  {Viennot}, \citenamefont {Dartiailh}, \citenamefont {Cottet},\ and\
  \citenamefont {Kontos}}]{viennot2015coherent}%
  \BibitemOpen
  \bibfield  {author} {\bibinfo {author} {\bibfnamefont {JJ}~\bibnamefont
  {Viennot}}, \bibinfo {author} {\bibfnamefont {MC}~\bibnamefont {Dartiailh}},
  \bibinfo {author} {\bibfnamefont {Audrey}\ \bibnamefont {Cottet}}, \ and\
  \bibinfo {author} {\bibfnamefont {Takis}\ \bibnamefont {Kontos}},\ }\bibfield
   {title} {\enquote {\bibinfo {title} {Coherent coupling of a single spin to
  microwave cavity photons},}\ }\href@noop {} {\bibfield  {journal} {\bibinfo
  {journal} {Science}\ }\textbf {\bibinfo {volume} {349}},\ \bibinfo {pages}
  {408--411} (\bibinfo {year} {2015})}\BibitemShut {NoStop}%
\bibitem [{\citenamefont {Guo}\ \emph {et~al.}(2018{\natexlab{a}})\citenamefont
  {Guo}, \citenamefont {Jiang}, \citenamefont {Li}, \citenamefont {Chen},\ and\
  \citenamefont {Agarwal}}]{guo2018enhancement}%
  \BibitemOpen
  \bibfield  {author} {\bibinfo {author} {\bibfnamefont {Zhiwei}\ \bibnamefont
  {Guo}}, \bibinfo {author} {\bibfnamefont {Haitao}\ \bibnamefont {Jiang}},
  \bibinfo {author} {\bibfnamefont {Yunhui}\ \bibnamefont {Li}}, \bibinfo
  {author} {\bibfnamefont {Hong}\ \bibnamefont {Chen}}, \ and\ \bibinfo
  {author} {\bibfnamefont {GS}~\bibnamefont {Agarwal}},\ }\bibfield  {title}
  {\enquote {\bibinfo {title} {Enhancement of electromagnetically induced
  transparency in metamaterials using long range coupling mediated by a
  hyperbolic material},}\ }\href@noop {} {\bibfield  {journal} {\bibinfo
  {journal} {Opt. Express}\ }\textbf {\bibinfo {volume} {26}},\ \bibinfo
  {pages} {627--641} (\bibinfo {year} {2018}{\natexlab{a}})}\BibitemShut
  {NoStop}%
\bibitem [{\citenamefont {Atabaki}\ \emph {et~al.}(2018)\citenamefont
  {Atabaki}, \citenamefont {Moazeni}, \citenamefont {Pavanello}, \citenamefont
  {Gevorgyan}, \citenamefont {Notaros}, \citenamefont {Alloatti}, \citenamefont
  {Wade}, \citenamefont {Sun}, \citenamefont {Kruger}, \citenamefont {Meng}
  \emph {et~al.}}]{atabaki2018integrating}%
  \BibitemOpen
  \bibfield  {author} {\bibinfo {author} {\bibfnamefont {Amir~H}\ \bibnamefont
  {Atabaki}}, \bibinfo {author} {\bibfnamefont {Sajjad}\ \bibnamefont
  {Moazeni}}, \bibinfo {author} {\bibfnamefont {Fabio}\ \bibnamefont
  {Pavanello}}, \bibinfo {author} {\bibfnamefont {Hayk}\ \bibnamefont
  {Gevorgyan}}, \bibinfo {author} {\bibfnamefont {Jelena}\ \bibnamefont
  {Notaros}}, \bibinfo {author} {\bibfnamefont {Luca}\ \bibnamefont
  {Alloatti}}, \bibinfo {author} {\bibfnamefont {Mark~T}\ \bibnamefont {Wade}},
  \bibinfo {author} {\bibfnamefont {Chen}\ \bibnamefont {Sun}}, \bibinfo
  {author} {\bibfnamefont {Seth~A}\ \bibnamefont {Kruger}}, \bibinfo {author}
  {\bibfnamefont {Huaiyu}\ \bibnamefont {Meng}},  \emph {et~al.},\ }\bibfield
  {title} {\enquote {\bibinfo {title} {Integrating photonics with silicon
  nanoelectronics for the next generation of systems on a chip},}\ }\href@noop
  {} {\bibfield  {journal} {\bibinfo  {journal} {Nature}\ }\textbf {\bibinfo
  {volume} {556}},\ \bibinfo {pages} {349} (\bibinfo {year}
  {2018})}\BibitemShut {NoStop}%
\bibitem [{\citenamefont {Bliokh}\ \emph
  {et~al.}(2015{\natexlab{a}})\citenamefont {Bliokh}, \citenamefont
  {Smirnova},\ and\ \citenamefont {Nori}}]{bliokh2015quantum}%
  \BibitemOpen
  \bibfield  {author} {\bibinfo {author} {\bibfnamefont {Konstantin~Y}\
  \bibnamefont {Bliokh}}, \bibinfo {author} {\bibfnamefont {Daria}\
  \bibnamefont {Smirnova}}, \ and\ \bibinfo {author} {\bibfnamefont {Franco}\
  \bibnamefont {Nori}},\ }\bibfield  {title} {\enquote {\bibinfo {title}
  {Quantum spin hall effect of light},}\ }\href@noop {} {\bibfield  {journal}
  {\bibinfo  {journal} {Science}\ }\textbf {\bibinfo {volume} {348}},\ \bibinfo
  {pages} {1448--1451} (\bibinfo {year} {2015}{\natexlab{a}})}\BibitemShut
  {NoStop}%
\bibitem [{\citenamefont {Lodahl}\ \emph {et~al.}(2017)\citenamefont {Lodahl},
  \citenamefont {Mahmoodian}, \citenamefont {Stobbe}, \citenamefont
  {Rauschenbeutel}, \citenamefont {Schneeweiss}, \citenamefont {Volz},
  \citenamefont {Pichler},\ and\ \citenamefont {Zoller}}]{lodahl2017chiral}%
  \BibitemOpen
  \bibfield  {author} {\bibinfo {author} {\bibfnamefont {Peter}\ \bibnamefont
  {Lodahl}}, \bibinfo {author} {\bibfnamefont {Sahand}\ \bibnamefont
  {Mahmoodian}}, \bibinfo {author} {\bibfnamefont {S{\o}ren}\ \bibnamefont
  {Stobbe}}, \bibinfo {author} {\bibfnamefont {Arno}\ \bibnamefont
  {Rauschenbeutel}}, \bibinfo {author} {\bibfnamefont {Philipp}\ \bibnamefont
  {Schneeweiss}}, \bibinfo {author} {\bibfnamefont {J{\"u}rgen}\ \bibnamefont
  {Volz}}, \bibinfo {author} {\bibfnamefont {Hannes}\ \bibnamefont {Pichler}},
  \ and\ \bibinfo {author} {\bibfnamefont {Peter}\ \bibnamefont {Zoller}},\
  }\bibfield  {title} {\enquote {\bibinfo {title} {Chiral quantum optics},}\
  }\href@noop {} {\bibfield  {journal} {\bibinfo  {journal} {Nature}\ }\textbf
  {\bibinfo {volume} {541}},\ \bibinfo {pages} {473--480} (\bibinfo {year}
  {2017})}\BibitemShut {NoStop}%
\bibitem [{\citenamefont {Aiello}\ \emph {et~al.}(2015)\citenamefont {Aiello},
  \citenamefont {Banzer}, \citenamefont {Neugebauer},\ and\ \citenamefont
  {Leuchs}}]{aiello2015transverse}%
  \BibitemOpen
  \bibfield  {author} {\bibinfo {author} {\bibfnamefont {Andrea}\ \bibnamefont
  {Aiello}}, \bibinfo {author} {\bibfnamefont {Peter}\ \bibnamefont {Banzer}},
  \bibinfo {author} {\bibfnamefont {Martin}\ \bibnamefont {Neugebauer}}, \ and\
  \bibinfo {author} {\bibfnamefont {Gerd}\ \bibnamefont {Leuchs}},\ }\bibfield
  {title} {\enquote {\bibinfo {title} {From transverse angular momentum to
  photonic wheels},}\ }\href@noop {} {\bibfield  {journal} {\bibinfo  {journal}
  {Nat. Photonics}\ }\textbf {\bibinfo {volume} {9}},\ \bibinfo {pages} {789}
  (\bibinfo {year} {2015})}\BibitemShut {NoStop}%
\bibitem [{\citenamefont {Bliokh}\ \emph
  {et~al.}(2015{\natexlab{b}})\citenamefont {Bliokh}, \citenamefont
  {Rodr{\'\i}guez-Fortu{\~n}o}, \citenamefont {Nori},\ and\ \citenamefont
  {Zayats}}]{bliokh2015spin}%
  \BibitemOpen
  \bibfield  {author} {\bibinfo {author} {\bibfnamefont {Konstantin~Yu}\
  \bibnamefont {Bliokh}}, \bibinfo {author} {\bibfnamefont {Francisco~J}\
  \bibnamefont {Rodr{\'\i}guez-Fortu{\~n}o}}, \bibinfo {author} {\bibfnamefont
  {Franco}\ \bibnamefont {Nori}}, \ and\ \bibinfo {author} {\bibfnamefont
  {Anatoly~V}\ \bibnamefont {Zayats}},\ }\bibfield  {title} {\enquote {\bibinfo
  {title} {Spin--orbit interactions of light},}\ }\href@noop {} {\bibfield
  {journal} {\bibinfo  {journal} {Nat. Photonics}\ }\textbf {\bibinfo {volume}
  {9}},\ \bibinfo {pages} {796} (\bibinfo {year}
  {2015}{\natexlab{b}})}\BibitemShut {NoStop}%
\bibitem [{\citenamefont {S{\"o}llner}\ \emph {et~al.}(2015)\citenamefont
  {S{\"o}llner}, \citenamefont {Mahmoodian}, \citenamefont {Hansen},
  \citenamefont {Midolo}, \citenamefont {Javadi}, \citenamefont
  {Kir{\v{s}}ansk{\.e}}, \citenamefont {Pregnolato}, \citenamefont {El-Ella},
  \citenamefont {Lee}, \citenamefont {Song} \emph
  {et~al.}}]{sollner2015deterministic}%
  \BibitemOpen
  \bibfield  {author} {\bibinfo {author} {\bibfnamefont {Immo}\ \bibnamefont
  {S{\"o}llner}}, \bibinfo {author} {\bibfnamefont {Sahand}\ \bibnamefont
  {Mahmoodian}}, \bibinfo {author} {\bibfnamefont {Sofie~Lindskov}\
  \bibnamefont {Hansen}}, \bibinfo {author} {\bibfnamefont {Leonardo}\
  \bibnamefont {Midolo}}, \bibinfo {author} {\bibfnamefont {Alisa}\
  \bibnamefont {Javadi}}, \bibinfo {author} {\bibfnamefont {Gabija}\
  \bibnamefont {Kir{\v{s}}ansk{\.e}}}, \bibinfo {author} {\bibfnamefont
  {Tommaso}\ \bibnamefont {Pregnolato}}, \bibinfo {author} {\bibfnamefont
  {Haitham}\ \bibnamefont {El-Ella}}, \bibinfo {author} {\bibfnamefont
  {Eun~Hye}\ \bibnamefont {Lee}}, \bibinfo {author} {\bibfnamefont {Jin~Dong}\
  \bibnamefont {Song}},  \emph {et~al.},\ }\bibfield  {title} {\enquote
  {\bibinfo {title} {Deterministic photon--emitter coupling in chiral photonic
  circuits},}\ }\href@noop {} {\bibfield  {journal} {\bibinfo  {journal} {Nat.
  Nanotech.}\ }\textbf {\bibinfo {volume} {10}},\ \bibinfo {pages} {775}
  (\bibinfo {year} {2015})}\BibitemShut {NoStop}%
\bibitem [{\citenamefont {Rodr{\'\i}guez-Fortu{\~n}o}\ \emph
  {et~al.}(2013)\citenamefont {Rodr{\'\i}guez-Fortu{\~n}o}, \citenamefont
  {Marino}, \citenamefont {Ginzburg}, \citenamefont {O’Connor}, \citenamefont
  {Mart{\'\i}nez}, \citenamefont {Wurtz},\ and\ \citenamefont
  {Zayats}}]{rodriguez2013near}%
  \BibitemOpen
  \bibfield  {author} {\bibinfo {author} {\bibfnamefont {Francisco~J}\
  \bibnamefont {Rodr{\'\i}guez-Fortu{\~n}o}}, \bibinfo {author} {\bibfnamefont
  {Giuseppe}\ \bibnamefont {Marino}}, \bibinfo {author} {\bibfnamefont {Pavel}\
  \bibnamefont {Ginzburg}}, \bibinfo {author} {\bibfnamefont {Daniel}\
  \bibnamefont {O’Connor}}, \bibinfo {author} {\bibfnamefont {Alejandro}\
  \bibnamefont {Mart{\'\i}nez}}, \bibinfo {author} {\bibfnamefont {Gregory~A}\
  \bibnamefont {Wurtz}}, \ and\ \bibinfo {author} {\bibfnamefont {Anatoly~V}\
  \bibnamefont {Zayats}},\ }\bibfield  {title} {\enquote {\bibinfo {title}
  {Near-field interference for the unidirectional excitation of electromagnetic
  guided modes},}\ }\href@noop {} {\bibfield  {journal} {\bibinfo  {journal}
  {Science}\ }\textbf {\bibinfo {volume} {340}},\ \bibinfo {pages} {328--330}
  (\bibinfo {year} {2013})}\BibitemShut {NoStop}%
\bibitem [{\citenamefont {Van~Mechelen}\ and\ \citenamefont
  {Jacob}(2016)}]{van2016universal}%
  \BibitemOpen
  \bibfield  {author} {\bibinfo {author} {\bibfnamefont {Todd}\ \bibnamefont
  {Van~Mechelen}}\ and\ \bibinfo {author} {\bibfnamefont {Zubin}\ \bibnamefont
  {Jacob}},\ }\bibfield  {title} {\enquote {\bibinfo {title} {Universal
  spin-momentum locking of evanescent waves},}\ }\href@noop {} {\bibfield
  {journal} {\bibinfo  {journal} {Optica}\ }\textbf {\bibinfo {volume} {3}},\
  \bibinfo {pages} {118--126} (\bibinfo {year} {2016})}\BibitemShut {NoStop}%
\bibitem [{\citenamefont {Shi}\ \emph {et~al.}(2019)\citenamefont {Shi},
  \citenamefont {Zhao}, \citenamefont {Long}, \citenamefont {Yang},
  \citenamefont {Wang}, \citenamefont {Chen}, \citenamefont {Ren},\ and\
  \citenamefont {Zhang}}]{shi2019observation}%
  \BibitemOpen
  \bibfield  {author} {\bibinfo {author} {\bibfnamefont {Chengzhi}\
  \bibnamefont {Shi}}, \bibinfo {author} {\bibfnamefont {Rongkuo}\ \bibnamefont
  {Zhao}}, \bibinfo {author} {\bibfnamefont {Yang}\ \bibnamefont {Long}},
  \bibinfo {author} {\bibfnamefont {Sui}\ \bibnamefont {Yang}}, \bibinfo
  {author} {\bibfnamefont {Yuan}\ \bibnamefont {Wang}}, \bibinfo {author}
  {\bibfnamefont {Hong}\ \bibnamefont {Chen}}, \bibinfo {author} {\bibfnamefont
  {Jie}\ \bibnamefont {Ren}}, \ and\ \bibinfo {author} {\bibfnamefont {Xiang}\
  \bibnamefont {Zhang}},\ }\bibfield  {title} {\enquote {\bibinfo {title}
  {Observation of acoustic spin},}\ }\href@noop {} {\bibfield  {journal}
  {\bibinfo  {journal} {Natl. Sci. Rev.}\ }\textbf {\bibinfo {volume} {6}},\
  \bibinfo {pages} {707--712} (\bibinfo {year} {2019})}\BibitemShut {NoStop}%
\bibitem [{\citenamefont {Long}\ \emph {et~al.}(2018)\citenamefont {Long},
  \citenamefont {Ren},\ and\ \citenamefont {Chen}}]{long2018intrinsic}%
  \BibitemOpen
  \bibfield  {author} {\bibinfo {author} {\bibfnamefont {Yang}\ \bibnamefont
  {Long}}, \bibinfo {author} {\bibfnamefont {Jie}\ \bibnamefont {Ren}}, \ and\
  \bibinfo {author} {\bibfnamefont {Hong}\ \bibnamefont {Chen}},\ }\bibfield
  {title} {\enquote {\bibinfo {title} {Intrinsic spin of elastic waves},}\
  }\href@noop {} {\bibfield  {journal} {\bibinfo  {journal} {Proc. Natl. Acad.
  Sci. U.S.A.}\ }\textbf {\bibinfo {volume} {115}},\ \bibinfo {pages}
  {9951--9955} (\bibinfo {year} {2018})}\BibitemShut {NoStop}%
\bibitem [{\citenamefont {Picardi}\ \emph {et~al.}(2018)\citenamefont
  {Picardi}, \citenamefont {Zayats},\ and\ \citenamefont
  {Rodr{\'\i}guez-Fortu{\~n}o}}]{picardi2018janus}%
  \BibitemOpen
  \bibfield  {author} {\bibinfo {author} {\bibfnamefont {Michela~F}\
  \bibnamefont {Picardi}}, \bibinfo {author} {\bibfnamefont {Anatoly~V}\
  \bibnamefont {Zayats}}, \ and\ \bibinfo {author} {\bibfnamefont
  {Francisco~J}\ \bibnamefont {Rodr{\'\i}guez-Fortu{\~n}o}},\ }\bibfield
  {title} {\enquote {\bibinfo {title} {Janus and huygens dipoles: near-field
  directionality beyond spin-momentum locking},}\ }\href@noop {} {\bibfield
  {journal} {\bibinfo  {journal} {Phys. Rev. Lett.}\ }\textbf {\bibinfo
  {volume} {120}},\ \bibinfo {pages} {117402} (\bibinfo {year}
  {2018})}\BibitemShut {NoStop}%
\bibitem [{\citenamefont {Picardi}\ \emph {et~al.}(2019)\citenamefont
  {Picardi}, \citenamefont {Neugebauer}, \citenamefont {Eismann}, \citenamefont
  {Leuchs}, \citenamefont {Banzer}, \citenamefont
  {Rodr{\'\i}guez-Fortu{\~n}o},\ and\ \citenamefont
  {Zayats}}]{picardi2019experimental}%
  \BibitemOpen
  \bibfield  {author} {\bibinfo {author} {\bibfnamefont {Michela~F}\
  \bibnamefont {Picardi}}, \bibinfo {author} {\bibfnamefont {Martin}\
  \bibnamefont {Neugebauer}}, \bibinfo {author} {\bibfnamefont {J{\"o}rg~S}\
  \bibnamefont {Eismann}}, \bibinfo {author} {\bibfnamefont {Gerd}\
  \bibnamefont {Leuchs}}, \bibinfo {author} {\bibfnamefont {Peter}\
  \bibnamefont {Banzer}}, \bibinfo {author} {\bibfnamefont {Francisco~J}\
  \bibnamefont {Rodr{\'\i}guez-Fortu{\~n}o}}, \ and\ \bibinfo {author}
  {\bibfnamefont {Anatoly~V}\ \bibnamefont {Zayats}},\ }\bibfield  {title}
  {\enquote {\bibinfo {title} {Experimental demonstration of linear and
  spinning janus dipoles for polarisation-and wavelength-selective near-field
  coupling},}\ }\href@noop {} {\bibfield  {journal} {\bibinfo  {journal} {Light
  Sci. Appl.}\ }\textbf {\bibinfo {volume} {8}},\ \bibinfo {pages} {52}
  (\bibinfo {year} {2019})}\BibitemShut {NoStop}%
\bibitem [{\citenamefont {F.~Picardi}\ \emph {et~al.}(2019)\citenamefont
  {F.~Picardi}, \citenamefont {V.~Zayats},\ and\ \citenamefont
  {J.~Rodríguez-Fortuño}}]{F.Picardi2019}%
  \BibitemOpen
  \bibfield  {author} {\bibinfo {author} {\bibfnamefont {Michela}\ \bibnamefont
  {F.~Picardi}}, \bibinfo {author} {\bibfnamefont {Anatoly}\ \bibnamefont
  {V.~Zayats}}, \ and\ \bibinfo {author} {\bibfnamefont {Francisco}\
  \bibnamefont {J.~Rodríguez-Fortuño}},\ }\bibfield  {title} {\enquote
  {\bibinfo {title} {Amplitude and phase control of guided modes excitation
  from a single dipole source: Engineering far- and near-field
  directionality},}\ }\href@noop {} {\bibfield  {journal} {\bibinfo  {journal}
  {Laser Photonics Rev.}\ }\textbf {\bibinfo {volume} {13}},\ \bibinfo {pages}
  {1900250} (\bibinfo {year} {2019})}\BibitemShut {NoStop}%
\bibitem [{\citenamefont {Guo}\ \emph {et~al.}(2017)\citenamefont {Guo},
  \citenamefont {Jiang}, \citenamefont {Long}, \citenamefont {Yu},
  \citenamefont {Ren}, \citenamefont {Xue},\ and\ \citenamefont
  {Chen}}]{guo2017photonic}%
  \BibitemOpen
  \bibfield  {author} {\bibinfo {author} {\bibfnamefont {Zhiwei}\ \bibnamefont
  {Guo}}, \bibinfo {author} {\bibfnamefont {Haitao}\ \bibnamefont {Jiang}},
  \bibinfo {author} {\bibfnamefont {Yang}\ \bibnamefont {Long}}, \bibinfo
  {author} {\bibfnamefont {Kun}\ \bibnamefont {Yu}}, \bibinfo {author}
  {\bibfnamefont {Jie}\ \bibnamefont {Ren}}, \bibinfo {author} {\bibfnamefont
  {Chunhua}\ \bibnamefont {Xue}}, \ and\ \bibinfo {author} {\bibfnamefont
  {Hong}\ \bibnamefont {Chen}},\ }\bibfield  {title} {\enquote {\bibinfo
  {title} {Photonic spin hall effect in waveguides composed of two types of
  single-negative metamaterials},}\ }\href@noop {} {\bibfield  {journal}
  {\bibinfo  {journal} {Sci. Rep.}\ }\textbf {\bibinfo {volume} {7}},\ \bibinfo
  {pages} {7742} (\bibinfo {year} {2017})}\BibitemShut {NoStop}%
\bibitem [{\citenamefont {Huard}\ and\ \citenamefont
  {Imbert}(1978)}]{huard1978measurement}%
  \BibitemOpen
  \bibfield  {author} {\bibinfo {author} {\bibfnamefont {S}~\bibnamefont
  {Huard}}\ and\ \bibinfo {author} {\bibfnamefont {Ch}~\bibnamefont {Imbert}},\
  }\bibfield  {title} {\enquote {\bibinfo {title} {Measurement of exchanged
  momentum during interaction between surface-wave and moving atom},}\
  }\href@noop {} {\bibfield  {journal} {\bibinfo  {journal} {Optics
  Communications}\ }\textbf {\bibinfo {volume} {24}},\ \bibinfo {pages}
  {185--189} (\bibinfo {year} {1978})}\BibitemShut {NoStop}%
\bibitem [{\citenamefont {Matsudo}\ \emph {et~al.}(1998)\citenamefont
  {Matsudo}, \citenamefont {Takahara}, \citenamefont {Hori},\ and\
  \citenamefont {Sakurai}}]{matsudo1998pseudomomentum}%
  \BibitemOpen
  \bibfield  {author} {\bibinfo {author} {\bibfnamefont {Tatsuo}\ \bibnamefont
  {Matsudo}}, \bibinfo {author} {\bibfnamefont {Y{\=u}ichir{\=o}}\ \bibnamefont
  {Takahara}}, \bibinfo {author} {\bibfnamefont {Hirokazu}\ \bibnamefont
  {Hori}}, \ and\ \bibinfo {author} {\bibfnamefont {Takeki}\ \bibnamefont
  {Sakurai}},\ }\bibfield  {title} {\enquote {\bibinfo {title} {Pseudomomentum
  transfer from evanescent waves to atoms measured by saturated absorption
  spectroscopy},}\ }\href@noop {} {\bibfield  {journal} {\bibinfo  {journal}
  {Optics communications}\ }\textbf {\bibinfo {volume} {145}},\ \bibinfo
  {pages} {64--68} (\bibinfo {year} {1998})}\BibitemShut {NoStop}%
\bibitem [{\citenamefont {Bliokh}\ \emph {et~al.}(2019)\citenamefont {Bliokh},
  \citenamefont {Leykam}, \citenamefont {Lein},\ and\ \citenamefont
  {Nori}}]{bliokh2019topological}%
  \BibitemOpen
  \bibfield  {author} {\bibinfo {author} {\bibfnamefont {Konstantin~Y}\
  \bibnamefont {Bliokh}}, \bibinfo {author} {\bibfnamefont {Daniel}\
  \bibnamefont {Leykam}}, \bibinfo {author} {\bibfnamefont {Max}\ \bibnamefont
  {Lein}}, \ and\ \bibinfo {author} {\bibfnamefont {Franco}\ \bibnamefont
  {Nori}},\ }\bibfield  {title} {\enquote {\bibinfo {title} {Topological
  non-hermitian origin of surface maxwell waves},}\ }\href@noop {} {\bibfield
  {journal} {\bibinfo  {journal} {Nat. Commun.}\ }\textbf {\bibinfo {volume}
  {10}},\ \bibinfo {pages} {580} (\bibinfo {year} {2019})}\BibitemShut
  {NoStop}%
\bibitem [{\citenamefont {Balanis}(1999)}]{balanis1999advanced}%
  \BibitemOpen
  \bibfield  {author} {\bibinfo {author} {\bibfnamefont {Constantine~A}\
  \bibnamefont {Balanis}},\ }\href@noop {} {\emph {\bibinfo {title} {Advanced
  engineering electromagnetics}}}\ (\bibinfo  {publisher} {John Wiley \&
  Sons},\ \bibinfo {year} {1999})\BibitemShut {NoStop}%
\bibitem [{\citenamefont {Picardi}\ \emph {et~al.}(2017)\citenamefont
  {Picardi}, \citenamefont {Manjavacas}, \citenamefont {Zayats},\ and\
  \citenamefont {Rodr{\'\i}guez-Fortu{\~n}o}}]{picardi2017unidirectional}%
  \BibitemOpen
  \bibfield  {author} {\bibinfo {author} {\bibfnamefont {Michela~F}\
  \bibnamefont {Picardi}}, \bibinfo {author} {\bibfnamefont {Alejandro}\
  \bibnamefont {Manjavacas}}, \bibinfo {author} {\bibfnamefont {Anatoly~V}\
  \bibnamefont {Zayats}}, \ and\ \bibinfo {author} {\bibfnamefont
  {Francisco~J}\ \bibnamefont {Rodr{\'\i}guez-Fortu{\~n}o}},\ }\bibfield
  {title} {\enquote {\bibinfo {title} {Unidirectional evanescent-wave coupling
  from circularly polarized electric and magnetic dipoles: An angular spectrum
  approach},}\ }\href@noop {} {\bibfield  {journal} {\bibinfo  {journal} {Phys.
  Rev. B}\ }\textbf {\bibinfo {volume} {95}},\ \bibinfo {pages} {245416}
  (\bibinfo {year} {2017})}\BibitemShut {NoStop}%
\bibitem [{\citenamefont {V\'azquez-Lozano}\ \emph {et~al.}(2019)\citenamefont
  {V\'azquez-Lozano}, \citenamefont {Mart\'{\i}nez},\ and\ \citenamefont
  {Rodr\'{\i}guez-Fortu\~no}}]{PhysRevApplied.12.024065}%
  \BibitemOpen
  \bibfield  {author} {\bibinfo {author} {\bibfnamefont {J.~Enrique}\
  \bibnamefont {V\'azquez-Lozano}}, \bibinfo {author} {\bibfnamefont
  {Alejandro}\ \bibnamefont {Mart\'{\i}nez}}, \ and\ \bibinfo {author}
  {\bibfnamefont {Francisco~J.}\ \bibnamefont {Rodr\'{\i}guez-Fortu\~no}},\
  }\bibfield  {title} {\enquote {\bibinfo {title} {Near-field directionality
  beyond the dipole approximation: Electric quadrupole and higher-order
  multipole angular spectra},}\ }\href@noop {} {\bibfield  {journal} {\bibinfo
  {journal} {Phys. Rev. Applied}\ }\textbf {\bibinfo {volume} {12}},\ \bibinfo
  {pages} {024065} (\bibinfo {year} {2019})}\BibitemShut {NoStop}%
\bibitem [{\citenamefont {Ishimaru}(2017)}]{ishimaru2017electromagnetic}%
  \BibitemOpen
  \bibfield  {author} {\bibinfo {author} {\bibfnamefont {Akira}\ \bibnamefont
  {Ishimaru}},\ }\href@noop {} {\emph {\bibinfo {title} {Electromagnetic wave
  propagation, radiation, and scattering from fundamentals to applications}}}\
  (\bibinfo  {publisher} {Wiley Online Library},\ \bibinfo {year}
  {2017})\BibitemShut {NoStop}%
\bibitem [{\citenamefont {Guo}\ \emph {et~al.}(2018{\natexlab{b}})\citenamefont
  {Guo}, \citenamefont {Jiang}, \citenamefont {Zhu}, \citenamefont {Sun},
  \citenamefont {Li},\ and\ \citenamefont {Chen}}]{PhysRevApplied.10.064048}%
  \BibitemOpen
  \bibfield  {author} {\bibinfo {author} {\bibfnamefont {Zhiwei}\ \bibnamefont
  {Guo}}, \bibinfo {author} {\bibfnamefont {Haitao}\ \bibnamefont {Jiang}},
  \bibinfo {author} {\bibfnamefont {Kejia}\ \bibnamefont {Zhu}}, \bibinfo
  {author} {\bibfnamefont {Yong}\ \bibnamefont {Sun}}, \bibinfo {author}
  {\bibfnamefont {Yunhui}\ \bibnamefont {Li}}, \ and\ \bibinfo {author}
  {\bibfnamefont {Hong}\ \bibnamefont {Chen}},\ }\bibfield  {title} {\enquote
  {\bibinfo {title} {Focusing and super-resolution with partial cloaking based
  on linear-crossing metamaterials},}\ }\href@noop {} {\bibfield  {journal}
  {\bibinfo  {journal} {Phys. Rev. Applied}\ }\textbf {\bibinfo {volume}
  {10}},\ \bibinfo {pages} {064048} (\bibinfo {year}
  {2018}{\natexlab{b}})}\BibitemShut {NoStop}%
\bibitem [{\citenamefont {Kapitanova}\ \emph {et~al.}(2014)\citenamefont
  {Kapitanova}, \citenamefont {Ginzburg}, \citenamefont
  {Rodr{\'\i}guez-Fortu{\~n}o}, \citenamefont {Filonov}, \citenamefont
  {Voroshilov}, \citenamefont {Belov}, \citenamefont {Poddubny}, \citenamefont
  {Kivshar}, \citenamefont {Wurtz},\ and\ \citenamefont
  {Zayats}}]{kapitanova2014photonic}%
  \BibitemOpen
  \bibfield  {author} {\bibinfo {author} {\bibfnamefont {Polina~V}\
  \bibnamefont {Kapitanova}}, \bibinfo {author} {\bibfnamefont {Pavel}\
  \bibnamefont {Ginzburg}}, \bibinfo {author} {\bibfnamefont {Francisco~J}\
  \bibnamefont {Rodr{\'\i}guez-Fortu{\~n}o}}, \bibinfo {author} {\bibfnamefont
  {Dmitry~S}\ \bibnamefont {Filonov}}, \bibinfo {author} {\bibfnamefont
  {Pavel~M}\ \bibnamefont {Voroshilov}}, \bibinfo {author} {\bibfnamefont
  {Pavel~A}\ \bibnamefont {Belov}}, \bibinfo {author} {\bibfnamefont
  {Alexander~N}\ \bibnamefont {Poddubny}}, \bibinfo {author} {\bibfnamefont
  {Yuri~S}\ \bibnamefont {Kivshar}}, \bibinfo {author} {\bibfnamefont
  {Gregory~A}\ \bibnamefont {Wurtz}}, \ and\ \bibinfo {author} {\bibfnamefont
  {Anatoly~V}\ \bibnamefont {Zayats}},\ }\bibfield  {title} {\enquote {\bibinfo
  {title} {Photonic spin hall effect in hyperbolic metamaterials for
  polarization-controlled routing of subwavelength modes},}\ }\href@noop {}
  {\bibfield  {journal} {\bibinfo  {journal} {Nat. Commun.}\ }\textbf {\bibinfo
  {volume} {5}},\ \bibinfo {pages} {3226} (\bibinfo {year} {2014})}\BibitemShut
  {NoStop}%
\bibitem [{\citenamefont {Li}\ \emph {et~al.}(2018)\citenamefont {Li},
  \citenamefont {Sun}, \citenamefont {Zhu}, \citenamefont {Guo}, \citenamefont
  {Jiang}, \citenamefont {Kariyado}, \citenamefont {Chen},\ and\ \citenamefont
  {Hu}}]{li2018topological}%
  \BibitemOpen
  \bibfield  {author} {\bibinfo {author} {\bibfnamefont {Yuan}\ \bibnamefont
  {Li}}, \bibinfo {author} {\bibfnamefont {Yong}\ \bibnamefont {Sun}}, \bibinfo
  {author} {\bibfnamefont {Weiwei}\ \bibnamefont {Zhu}}, \bibinfo {author}
  {\bibfnamefont {Zhiwei}\ \bibnamefont {Guo}}, \bibinfo {author}
  {\bibfnamefont {Jun}\ \bibnamefont {Jiang}}, \bibinfo {author} {\bibfnamefont
  {Toshikaze}\ \bibnamefont {Kariyado}}, \bibinfo {author} {\bibfnamefont
  {Hong}\ \bibnamefont {Chen}}, \ and\ \bibinfo {author} {\bibfnamefont {Xiao}\
  \bibnamefont {Hu}},\ }\bibfield  {title} {\enquote {\bibinfo {title}
  {Topological lc-circuits based on microstrips and observation of
  electromagnetic modes with orbital angular momentum},}\ }\href@noop {}
  {\bibfield  {journal} {\bibinfo  {journal} {Nat. Commun.}\ }\textbf {\bibinfo
  {volume} {9}},\ \bibinfo {pages} {4598} (\bibinfo {year} {2018})}\BibitemShut
  {NoStop}%
\bibitem [{\citenamefont {Assawaworrarit}\ \emph {et~al.}(2017)\citenamefont
  {Assawaworrarit}, \citenamefont {Yu},\ and\ \citenamefont
  {Fan}}]{assawaworrarit2017robust}%
  \BibitemOpen
  \bibfield  {author} {\bibinfo {author} {\bibfnamefont {Sid}\ \bibnamefont
  {Assawaworrarit}}, \bibinfo {author} {\bibfnamefont {Xiaofang}\ \bibnamefont
  {Yu}}, \ and\ \bibinfo {author} {\bibfnamefont {Shanhui}\ \bibnamefont
  {Fan}},\ }\bibfield  {title} {\enquote {\bibinfo {title} {Robust wireless
  power transfer using a nonlinear parity--time-symmetric circuit},}\
  }\href@noop {} {\bibfield  {journal} {\bibinfo  {journal} {Nature}\ }\textbf
  {\bibinfo {volume} {546}},\ \bibinfo {pages} {387} (\bibinfo {year}
  {2017})}\BibitemShut {NoStop}%
\bibitem [{\citenamefont {Kurs}\ \emph {et~al.}(2007)\citenamefont {Kurs},
  \citenamefont {Karalis}, \citenamefont {Moffatt}, \citenamefont
  {Joannopoulos}, \citenamefont {Fisher},\ and\ \citenamefont
  {Solja{\v{c}}i{\'c}}}]{kurs2007wireless}%
  \BibitemOpen
  \bibfield  {author} {\bibinfo {author} {\bibfnamefont {Andre}\ \bibnamefont
  {Kurs}}, \bibinfo {author} {\bibfnamefont {Aristeidis}\ \bibnamefont
  {Karalis}}, \bibinfo {author} {\bibfnamefont {Robert}\ \bibnamefont
  {Moffatt}}, \bibinfo {author} {\bibfnamefont {John~D}\ \bibnamefont
  {Joannopoulos}}, \bibinfo {author} {\bibfnamefont {Peter}\ \bibnamefont
  {Fisher}}, \ and\ \bibinfo {author} {\bibfnamefont {Marin}\ \bibnamefont
  {Solja{\v{c}}i{\'c}}},\ }\bibfield  {title} {\enquote {\bibinfo {title}
  {Wireless power transfer via strongly coupled magnetic resonances},}\
  }\href@noop {} {\bibfield  {journal} {\bibinfo  {journal} {Science}\ }\textbf
  {\bibinfo {volume} {317}},\ \bibinfo {pages} {83--86} (\bibinfo {year}
  {2007})}\BibitemShut {NoStop}%
\bibitem [{\citenamefont {Gong}\ \emph {et~al.}(2018)\citenamefont {Gong},
  \citenamefont {Alpeggiani}, \citenamefont {Sciacca}, \citenamefont
  {Garnett},\ and\ \citenamefont {Kuipers}}]{gong2018nanoscale}%
  \BibitemOpen
  \bibfield  {author} {\bibinfo {author} {\bibfnamefont {Su-Hyun}\ \bibnamefont
  {Gong}}, \bibinfo {author} {\bibfnamefont {Filippo}\ \bibnamefont
  {Alpeggiani}}, \bibinfo {author} {\bibfnamefont {Beniamino}\ \bibnamefont
  {Sciacca}}, \bibinfo {author} {\bibfnamefont {Erik~C}\ \bibnamefont
  {Garnett}}, \ and\ \bibinfo {author} {\bibfnamefont {L}~\bibnamefont
  {Kuipers}},\ }\bibfield  {title} {\enquote {\bibinfo {title} {Nanoscale
  chiral valley-photon interface through optical spin-orbit coupling},}\
  }\href@noop {} {\bibfield  {journal} {\bibinfo  {journal} {Science}\ }\textbf
  {\bibinfo {volume} {359}},\ \bibinfo {pages} {443--447} (\bibinfo {year}
  {2018})}\BibitemShut {NoStop}%
\bibitem [{\citenamefont {Neugebauer}\ \emph {et~al.}(2019)\citenamefont
  {Neugebauer}, \citenamefont {Banzer},\ and\ \citenamefont
  {Nechayev}}]{neugebauer2019emission}%
  \BibitemOpen
  \bibfield  {author} {\bibinfo {author} {\bibfnamefont {Martin}\ \bibnamefont
  {Neugebauer}}, \bibinfo {author} {\bibfnamefont {Peter}\ \bibnamefont
  {Banzer}}, \ and\ \bibinfo {author} {\bibfnamefont {Sergey}\ \bibnamefont
  {Nechayev}},\ }\bibfield  {title} {\enquote {\bibinfo {title} {Emission of
  circularly polarized light by a linear dipole},}\ }\href@noop {} {\bibfield
  {journal} {\bibinfo  {journal} {Sci. Adv.}\ }\textbf {\bibinfo {volume}
  {5}},\ \bibinfo {pages} {eaav7588} (\bibinfo {year} {2019})}\BibitemShut
  {NoStop}%
\bibitem [{\citenamefont {Wang}\ \emph {et~al.}(2016)\citenamefont {Wang},
  \citenamefont {Cai}, \citenamefont {Zhang}, \citenamefont {Xu},\ and\
  \citenamefont {Luo}}]{wang2016directional}%
  \BibitemOpen
  \bibfield  {author} {\bibinfo {author} {\bibfnamefont {Lei}\ \bibnamefont
  {Wang}}, \bibinfo {author} {\bibfnamefont {Wei}\ \bibnamefont {Cai}},
  \bibinfo {author} {\bibfnamefont {Xinzheng}\ \bibnamefont {Zhang}}, \bibinfo
  {author} {\bibfnamefont {Jingjun}\ \bibnamefont {Xu}}, \ and\ \bibinfo
  {author} {\bibfnamefont {Yongsong}\ \bibnamefont {Luo}},\ }\bibfield  {title}
  {\enquote {\bibinfo {title} {Directional generation of graphene plasmons by
  near field interference},}\ }\href@noop {} {\bibfield  {journal} {\bibinfo
  {journal} {Opt. Express}\ }\textbf {\bibinfo {volume} {24}},\ \bibinfo
  {pages} {19776--19787} (\bibinfo {year} {2016})}\BibitemShut {NoStop}%
\end{thebibliography}%

\end{document}